\documentclass[manuscript]{aastex}
\slugcomment{}

\shorttitle{The Distance to the $\sigma$~Ori cluster}
\shortauthors{Sherry et al.}

\begin{document}


\title{Main-Sequence Fitting Distance to the $\sigma$~Ori Cluster}


\author{W.\ H.\ Sherry}
\affil{National Solar Observatory, Tucson, AZ 85719}
\email{wsherry@nso.edu}

\author{F.\ M.\ Walter}
\affil{SUNY Stony Brook, Stony Brook, NY 11790}
\email{fwalter@astro.sunysb.edu}

\author{S.\ J.\ Wolk\altaffilmark{3}}
\email{swolk@cfa.harvard.edu}

\and

\author{N.\ R.\ Adams\altaffilmark{3}}
\email{nadams@head.cfa.harvard.edu}
\affil{Center for Astrophysics, 60 Garden Street, Cambridge, Cambridge, MA 02138}

%\affil{Center for Astrophysics, 60 Garden Street, Cambridge, Cambridge, MA 02138}


%\altaffiltext{1}{Visiting Astronomer, Cerro Tololo Inter-American Observatory.
%CTIO is operated by AURA, Inc.\ under contract to the National Science
%\altaffiltext{2}{Society of Fellows, Harvard University.}
%\altaffiltext{3}{present address: Center for Astrophysics,
%    60 Garden Street, Cambridge, MA 02138}
%\altaffiltext{4}{Visiting Programmer, Space Telescope Science Institute}
%\altaffiltext{5}{Patron, Alonso's Bar and Grill}


\begin{abstract}
The $\sigma$~Ori cluster is an unbound aggregate of a few hundred
 young, low--mass stars centered on the multiple system $\sigma$~Ori.
 This cluster is of great interest because it is at an age when
 roughly half of the stars have lost their protoplanetary disks, and
 the cluster has a very large population of brown dwarfs.  One of the
 largest sources of uncertainty in the properties of the cluster is
 that the distance is not well known.  The directly measured Hipparcos
 distance to $\sigma$~Ori~AB is 350$^{+120}_{-90}$~pc.  On the other
 hand, the distance to the Orion~OB1b subgroup (of which $\sigma$~Ori
 is thought to be a member), 473$\pm$40~pc, is far better determined,
 but it is an indirect estimate of the cluster's distance.  Also,
 Orion OB1b may have a depth of 40~pc along our line of sight.
%For an age of $\sim$2--3~Myrs, we find that the main %sequence
%turn--on is around B9 to A0.  
 We use main sequence fitting to 9 main sequence cluster members to
 estimate a best fit distance of 420$\pm30$~pc, assuming a metallicity
 of $-$0.16$\pm$0.11 or 444~pc assuming solar metallicity.  A distance
 as close as 350~pc is inconsistent with the observed brightnesses of
 the cluster members.  At the best fit distance, the age of the
 cluster is 2--3~Myrs.
% with spectral types earlier than F0
% that lie within 30$^{\prime}$ of $\sigma$~Ori to estimate the
% distance to the cluster.  
\end{abstract}





\section{Introduction}

%\citep{hen61,lyn68,spi85} \object{NGC 6397}, \citep[Paper I]{djo84}
%\objectname{NGC 6624}, \objectname[M 15]{NGC 7078}, and \object[Cl
%1938-341]{Terzan 8})

The bright O9.5V star \object{$\sigma$~Ori} is a trapezium--like
system with six known components.  The brightest component,
$\sigma$~Ori~AB (V=3.$^m$80), is a 0.25$^{{\prime}{\prime}}$ binary
\citep{horch01} with an O9V primary and a B0.5V secondary
\citep{edwards76} and an orbital period of $\sim$170~years or
$\sim$158~years \citep{heintz1974, heintz1997}.  The O9V primary,
$\sigma$~Ori~A, is itself a double lined spectroscopic
binary\footnote{We wish to thank Deane Peterson for sharing
unpublished results of his observations of $\sigma$~Ori~AB with us.}
\citep{peterson07, bolton74, micz1950}.  The spectral type of
$\sigma$~Ori~C is A2V.  The D and E components are B2V stars with
V$\simeq$6.8 and $\simeq$6.6 respectively (see
Table~\ref{true_table}).

The $\sigma$~Ori cluster was first recognized as a group of high--mass
stars by \citet{garrison67}, and as cluster of low--mass
pre-main-sequence stars by \citet{wws}.  Continuing work on the
cluster has revealed a young cluster of several hundred low--mass
stars \citep{paper1, burn05, kenyon05} and a rich population of brown
dwarfs \citep{bejar99,bejar01,bejar04, zapatero02}.  About one third
to one half of the stars retain their accretion disks
\citep{oliveira04, hern2007}.  The cluster is considered part of the
Orion OB1b association.  Age estimates for Orion OB1b range from less
than 2~Myrs \citep{brown94} up to 7~Myrs \citep{blaauw91}.  This is an
exceptionally interesting age because it is the age when
protoplanetary disks are making the transition from optically thick to
optically thin and may be the age when giant planets form.  The more
accurately the cluster age can be measured, the tighter the constraint
on disk lifetimes and the time available for giant planet formation
will be.  See \citet{handbook} for a recent review of observational
data on the low--mass population of the $\sigma$~Ori cluster.

The most significant source of uncertainty for the age of this cluster
is the uncertain distance to the cluster.  Many authors adopt the
Hipparcos distance of 350$^{+120}_{-90}$~pc for $\sigma$~Ori
\citep{hip} as the distance to the center of the cluster.  This has
the virtue of being a direct measurement of the distance to the
cluster, but it has an uncertainty of 30\%
%**************
(see \citet{schroeder04} for a discussion of biases on Hipparcos parallaxes of
O stars).  Others use the Hipparcos
distance to the Orion~OB1b subgroup (of which $\sigma$~Ori is a
member), 473$\pm$40~pc \citep{dezeeuw99}, as the distance to the
cluster.  This value is more precise because it is the average distance to 42
members of the association,   %*****************
yet it is only an indirect measurement of the distance to
$\sigma$~Ori.  Similarly, \citet{hern2005} find a distance of
443$\pm16$~pc from the Hipparcos parallaxes of the combined Orion~OB1b
and 1c subgroups.  The Orion~OBIb association has a size of
$\sim$30--40~pc across the sky, so it is likely to have a similar
depth along our line of sight.  The cluster could easily lie $>$20~pc in
front of or behind the center of Orion~OBIb.  %Recent radial velocity
%measurements indicate that the $\sigma$~Ori cluster is kinematically
%distinct from the average radial velocity of Orion OB1b \citep{jeffries06}.  
It would be preferable to have a direct measure of the cluster's
distance that is more precise than the Hipparcos distance to
$\sigma$~Ori.

%While reviewing published data on the distance to the $\sigma$~Ori
%cluster, we found
In a brief abstract, \citet{garrison67} said that main~sequence
fitting to 15 B stars near $\sigma$~Ori yielded a narrow main~sequence
at a distance modulus of 8.2 (440~pc).  Garrison did not correct for
the small values of reddening that some of the likely cluster members
have.  Garrison does not appear to have ever published a more detailed
description of this result.

In this paper we re--examine the main~sequence fitting distance for
the $\sigma$~Ori cluster using published spectroscopy and photometry
for the stars that lie within 30$^{\prime}$ of $\sigma$~Ori~AB and
have spectral types earlier than F0.

%\section{Data}

\section{Analysis and Data}\label{analysis}

We searched the literature\footnote{Our initial search relied heavily
 upon the SIMBAD data base.} for photometry and spectral types for all
 of the stars within 30$^{\prime}$ of $\sigma$~Ori~AB that have
 spectral types earlier than F0.  Several of the stars have B and V
 photoelectric photometry from multiple observations taken prior to 1980.
 Table~\ref{true_table} collects our adopted colors and magnitudes for
 the 19 stars we selected.

%****************************  new paragraph
We have also obtained new
spectra of those stars whose spectral types or colors
appeared discrepant. These spectra, obtained with the SMARTS/CTIO
1.5m RC spectrograph, have 1.6\AA\ resolution between about 3880 and 4500\AA.
Spectral types have been determined through visual comparison with a grid of
spectral standards obtained with the same same instrument.

%As a first step we compared the observed values with the zero age main
%sequence (ZAMS) \citep{turn76,turn79,sk82}.
%%The results are shown inFigure~\ref{raw_cmd}.  
%A distance of 440~pc appeared to be a reasonable fit, but some of the
%stars were quite a distance from the main sequence.  In particular,
%$\sigma$~Ori~AB did not fall near the main~sequence because it has a
%significant reddening, and it is a multiple system.  Several other
%discrepant stars appeared to have significant reddenings.

\subsection{Magnitudes and B$-$V Colors for Individual Stars}

Several of the stars listed in Table~\ref{true_table} can not be
directly compared to the main sequence because they are binaries or
have known problems with their photometric data.  Their photometry must be
corrected before being included in any estimate of the cluster
distance.

\subsubsection{$\sigma$~Ori~Aa, $\sigma$~Ori~Ab, and $\sigma$~Ori~B}

$\sigma$~Ori~A and $\sigma$~Ori~B are the two brightest cluster
members.  This visual pair, with a separation of
0.25$^{{\prime}{\prime}}$, lies roughly at the center of the cluster.
\citet{horch01} used speckle observations to derive a V band magnitude
difference of 1.$^m$25.  Similar results were found by \citet{tenb00}
who report $\Delta$V=1.$^m$24. Given a combined magnitude of
V=3.$^m$8 and an E(B$-$V) of $\sim$0.$^m$06 (see section \ref{red}), a
magnitude difference of 1.$^m$25 indicates that $\sigma$~Ori~A
%**************************************************
has V$_0\simeq$3.$^m$91 while $\sigma$~Ori~B
has V$_0\simeq$5.$^m$16.

%a and $\sigma$~Ori~Ab have a combined V$_0\simeq$3.$^m$91 while $\sigma$~Ori~B
%has a V$_0\simeq$5.$^m$16.

%******************
$\sigma$~Ori~A is itself a double-lined spectroscopic binary.
\citet{bolton74} estimated a $\Delta$V of $\sim$0.$^m$5 between
$\sigma$~Ori~Aa and $\sigma$~Ori~Ab.  This would require
$\sigma$~Ori~Aa and Ab to have V$_0$$\sim$4.$^m$4 and
$\sim$4.$^m$9, respectively.  There is no measured spectral type for
$\sigma$~Ori~Ab yet, so its UBV colors are unknown.  Assuming that it
is on the ZAMS, the spectral type a star that is 0.$^m$5 fainter than
an O9V star should B0V.

%This magnitude difference means that
%$\sigma$~Ori~B is about 30\% as bright as $\sigma$~Ori~A in the V
%band.  That requires that $\sigma$~Ori~A have an earlier spectral type
%than the combined spectral type of O9.5V and $\sigma$~Ori~B have a
%later spectral type.

\citet{edwards76} quotes spectral types of O9V and B0.5V for $\sigma$~Ori~A and
B, respectively.  Assuming an
uncertainty of $\pm$0.5 subtypes, we adopt values of
(B$-$V)$_0\simeq-$0.$^m$31$\pm$0.01 for $\sigma$~Ori~Aa and
(B$-$V)$_0\simeq-$0.$^m$28$\pm$0.02 for $\sigma$~Ori~B \citep{kh95}.

The observed B$-$V for $\sigma$~Ori~AB is $-$0.$^m$24.  Assuming an
intrinsic color (B$-$V)$_0$ = $-$0.$^m$30 for $\sigma$~Ori~AB ,
E(B$-$V) must be $\sim$0.$^m$06.  This is consistent with the observed
N(H) column density of 3.3$\times$10$^{20}$~cm$^{-2}$
\citep{frusc94,bohlin83}.  The uncertainty on the column density is
20\%.  This value is also consistent with published estimates of the
line of sight reddening to $\sigma$~Ori~AB (e.\ g. \citet{lee68}) and
with the N(H) column density of 3.6$\times$10$^{20}$~cm$^{-2}$ measured
by \citet{shull85}.

\subsubsection{$\sigma$~Ori~C}

\citet{gw58} measured the B and V magnitudes of $\sigma$~Ori~C.  They
noted that the observed B$-$V color of $\sigma$~Ori~C, $-$0.$^m$02, is
too blue for its spectral type, A2V, which should have
(B$-$V)$_0$=0.$^m$06 \citep{kh95}.  \citet{gw58} account for the
exceptionally blue color of $\sigma$~Ori~C as the result of scattered
light from $\sigma$~Ori~AB (11$^{{\prime}{\prime}}$ away) in the
aperture of the photometer.  The published V~magnitude is 8.$^m$79
\citep{gw58}, but that measurement was also contaminated by scattered
V~band light from $\sigma$~Ori~AB.  \citet{gw58} estimate that, after
correcting for scattered light, the true V~magnitude of $\sigma$~Ori~C
is $\sim$9.$^m$2.

%The true magnitude for $\sigma$~Ori~C can be calculated with two
%assumptions.  The shape of the PSF for $\sigma$~Ori~AB must not vary
%strongly in the wings between the B band and V band observations.  The
%shape of the wings of the PSF depends mainly upon the optics of the
%telescope.  For observations taken on the same night, we do not expect
%the PSF to vary much from one observation to the next or using
%different filters.  To calculate V for $\sigma$~Ori~C we also need to
%have a value for the B$-$V color that $\sigma$~Ori~C would have in the
%absence of scattered light from $\sigma$~Ori~AB.  If we use
%(B$-$V)$_0$=0.06, the unreddened color of an A2V star, we find a V 
%magnitude of 9.09.  If instead we assume that
%$\sigma$~Ori~C has the same E(B$-$V) as $\sigma$~Ori~AB, then we find
%that the reddened V magnitude should be 9.26.  This would give
%$\sigma$~Ori~C an unreddened magnitude of V=9.07.  Assuming an
%uncertainty of 1 subtype on the spectral class leads to an uncertainty
%of $\pm$0.03 magnitudes on the B$-$V color.  Considering both the
%uncertainties on the intrinsic color and reddening of $\sigma$~Ori~C,
%we estimate that it has V$_0$=9.1$\pm$0.1.  There is also a chance that
%the intrinsic colors are slightly different from the standard B$-$V
%color of an A2V star because it may be a rapid rotator \citep{gw58}.  Our
%estimate of an error of 0.03 magnitudes on the B$-$V color may be
%optimistic.

\citet{paper3} report recent differential V and I$_C$ photometry for
stars within 6$^{\prime}$ of $\sigma$~Ori~AB.  While $\sigma$~Ori~AB
is saturated, C, D, and E are not saturated.  They find that
$\sigma$~Ori~C is 2.$^m$63$\pm$0.01 magnitudes fainter than $\sigma$~Ori~D, and
2.$^m$74$\pm$0.01 magnitudes fainter than $\sigma$~Ori~E in the V band.  Using
the V magnitudes from Table~\ref{true_table} for $\sigma$~Ori~D and
$\sigma$~Ori~E yields V=9.$^m$44 and V=9.$^m$40 respectively for
$\sigma$~Ori~C.  These measurements were done using small apertures
which are not significantly contaminated by scattered light.
%The 0.15 mag.\ difference between
%the calibrations using $\sigma$~Ori~D and $\sigma$~Ori~E is probably
%due to the known variability of $\sigma$~Ori~E($\Delta$V$\sim$0.1~mag.).  See Section~\ref{sigE}.  
We will adopt V=9.$^m$42$\pm$0.02 for the magnitude of $\sigma$~Ori~C
(uncorrected for reddening).

%If one assumes a color that $\sigma$~Ori~C should have in the absence
%of scattered light from $\sigma$~Ori~AB and that the shape of wings of
%the PSF for $\sigma$~Ori~AB doesn't vary strongly between the B and V
%bands, the correct magnitude for $\sigma$~Ori~C can be calculated.

%If $\sigma$~Ori~C has the same amount of reddening as measured
%for $\sigma$~Ori~AB and $\sigma$~Ori~E, then the unreddened magnitude
%should be $\sim$9.0

\subsubsection{$\sigma$~Ori~D}

The three papers that report UBV photometry for $\sigma$~Ori~D quote
significantly different V magnitudes, and to a lesser
extent, colors.  \citet{gw58} report a V magnitude of 6.$^m$62.
Eighteen years later, \citet{vogt76} reported a V magnitude of
6.$^m$73.  \citet{guetter79} reported a V magnitude of 6.$^m$84.
These discrepant V magnitudes may indicate variability, or that stray
light from $\sigma$~Ori~AB affected the measurements of
$\sigma$~Ori~D.  \citet{ubv} list a weighted averaged of the UBV
photometry for $\sigma$~Ori~D which we have used in
Table~\ref{true_table}.

\subsubsection{$\sigma$~Ori~E}\label{sigE}

The B2Vp star $\sigma$~Ori~E has unusually strong He lines
\citep{gw58} which make it spectroscopically peculiar.  It has
variable line widths and photometric variations
($\Delta$mag$\sim$0.$^m$03--0.$^m$15) with a period of 1.19 days
\citep{hess76, town05}.  There are conflicting opinions as to whether
$\sigma$~Ori~E is physically associated with $\sigma$~Ori~AB.  Much of
the uncertainty surrounding the membership of $\sigma$~Ori~E with
$\sigma$~Ori~AB follows from uncertainty on the mass and evolutionary
status of $\sigma$~Ori~E.  \citet{gw58} estimated the absolute
magnitude of $\sigma$~Ori~E using three different methods, thereby
placing the star on or near the main sequence (which would put
$\sigma$~Ori~D, and $\sigma$~Ori~E at the same distance).  They found
that the equivalent widths of two components of the interstellar K
line are similar for both $\sigma$~Ori~AB and $\sigma$~Ori~E, as are
the radial velocities.  Attempts to model the UV flux from the V band
flux and spectroscopic features lead to models of $\sigma$~Ori~E that
have masses that are far too small ($\sim$3~M$_{\odot}$) for an early
B main sequence star (M$\sim$9 ~M$_{\odot}$).  The main reason for the
low masses in these models is that the profiles of the Balmer and
helium lines indicate a low gravity \citep{hunger89}.  More recent
models postulate emission from plasma clouds magnetically confined above
the photosphere \citep{town05} which may explain the discrepancy
between the gravity estimated from line profiles and data that
indicate that $\sigma$~Ori~E is a main sequence star.  Given the
significant uncertainties on the models, we take the observations
indicating that $\sigma$~Ori~E is a main sequence star with a normal
mass and radius for its spectral type at face value.

A B2V star has (B$-$V)$_0$ of $-$0.$^m$24 \citep{kh95}.  The measured B$-$V
for $\sigma$~Ori~E is $-$0.$^m$18 \citep{guetter79}, which makes E(B$-$V)
0.$^m$06.  This is consistent with the observed N(H) column density of
4.5$\times$10$^{20}$~cm$^{-2}$ \citep{frusc94, shull85}.  The
uncertainty on the column density is 20\%.

\subsubsection{BD~$-$02~1323C and HD~294272}\label{bd1323c}

BD~$-$02~1323C was not found by our initial SIMBAD search for early
type stars within 30$^{\prime}$ of $\sigma$~Ori.  This star came to
our attention because SIMBAD notes that HD~294272 is a member of a
triple system, ADS~4240 \citep{aitken32}.  ADS~4240A is HD~294271.
ADS~4240B is HD~294272 which is separated from HD~294271 by
$\sim$68$^{{\prime}{\prime}}$.  ADS~4240C (BD~$-$02~1323C) is
separated from HD~294272 by 8.5$^{{\prime}{\prime}}$.  SIMBAD listed a
V magnitude of 10.$^m$3 for BD~$-$02~1323C, and no other photometric
measurements.  This value is not correct.
%We don't know where this value came from.

\citet{guetter79} reported photometry for BD~$-$02~1323A and
BD~$-$02~1323B.  \citet{guetter81} used the same names when he
reported the spectral types.  SIMBAD, which does not recognize the
names BD~$-$02~1323A and BD~$-$02~1323B, assigned the
\citet{guetter79} photometry for BD~$-$02~1323B (ADS~4240C) to
BD~$-$02~1323 which is ADS~4240B or HD~294272.  Consequently,
HD~294272 (ADS~4240B) was listed in SIMBAD as having the photometry
and spectral type of BD~$-$02~1323C (ADS~4240C).  The measurements
for the two stars from \citet{guetter79, guetter81} and \citet{ubv}
have been correctly assigned in Table~\ref{true_table}.

%In the Aitken Double Star Catalog
%\citep{aitken32} HD~294271 is listed as ADS~4240A, HD~294272 is listed
%as ADS~4240B, and BD~$-$02~1323C is listed as ADS~4240C.  

%\citet{paper3} observed HD~294271, HD~294272, and BD~$-$02~1323C in
%the V and I bands.  We found that BD~$-$02~1323C is 0.28 magnitudes
%fainter than HD~294272 and 0.81 magnitudes fainter than HD~294271.
%Using the calibration from HD~294272 BD~$-$02~1323C would have a V
%magnitude of 9.05.  The calibration from HD~294271 would give
%BD~$-$02~1323C a V magnitude of 8.69.  We do not know why this result
%differs from the previously reported value of V=10.3.

%Since the data from \citet{paper3} were taken under non-photometric
%conditions, we prefer the calibration from HD~294272 which is only
%8.5$^{{\prime}{\prime}}$ from BD~$-$02~1323C.  We will adopt a value
%of V=9.05 for BD~$-$02~1323C.

%We could not find any published B magnitudes or B$-$V colors for
%BD~$-$02~1323C.  As part of an on-going project to reobserve early
%type members of Orion~OB1b, \citet{walter07} reports a spectral type
%of ?? for BD~$-$02~1323C.

\subsubsection{HD~294273 \& HD~294279}

We obtained new spectra for  HD~294273 \& HD~294279 since
the only published spectral types that we could find are
A2 and A3, respectively, in the HD catalog. The Ca~II~K lines are far
too strong for early A spectral types.
The revised spectral types are early F (F0-F3)
for HD~294279 and A7 for HD~294273.  We do not assign a
luminosity class, but it is likely that they are both class V.

%\subsection{Choosing A ZAMS}
%
%Much to our surprise, the largest obstical to determining the
%main-sequence fitting distance to the $\sigma$~Ori cluster was
%deciding whoose zero age main sequence (ZAMS) was best.  

\subsection{Reddening}\label{red}

We estimate the reddening of most of the stars in our sample by
comparing the observed B$-$V color to the (B$-$V)$_0$ expected for
each star's spectral type, and computing A$_V$ assuming R=3.1.
%For $\sigma$~Ori~AB and $\sigma$~Ori~E we
%also estimate the reddening from published estimates of N(H) along the
%line of sight.  
Column~8 of Table~\ref{true_table} lists these A$_V$ values for
probable cluster members.  The mean E(B$-$V) for probable cluster
members is 0.$^m$06$\pm$0.005 ($\sigma$~Ori~A and B were treated as a
single measurement).  The median E(B$-$V) is also 0.$^m$06.  All of
the probable cluster members have values of E(B$-$V) between 0.$^m$04
and 0.$^m$09, which is consistent with the 0.$^m$015 uncertainty on
E(B$-$V) for individual stars.  The mean E(B$-$V) of
0.$^m$06$\pm$0.005 makes the mean A$_V$ of the cluster
0.$^m$19$\pm$0.02, in agreement with the values quoted by
\citet{lee68, shull85} for $\sigma$~Ori~AB.

Assuming that E(U$-$B)=0.72E(B$-$V), we expect a mean
E(U$-$B)$\sim$0.$^m$04 mag.  This is comparable to or smaller than the
uncertainties on (U$-$B)$_0$ due to the change in (U$-$B)$_0$ from one
spectral type to the next along the ZAMS.  We found a median E(U$-$B)
of 0.$^m$02$\pm$0.03 with most of the stars in Table~\ref{true_table}
having values ranging from $-$0.$^m$06 to 0.$^m$07.  HD~37633 (V1147
Ori), a known variable, does have an exceptionally large, negative
E(U$-$B)=$-$0.$^m$16.  This may be due to its variability or a U
band excess.  Our spectrum shows a spectral type of B9.5.
%Our spectrum shows that HD~37633 has a peculiar mid--B spectrum around B6.  
%An earlier type could also account for the star's
%very blue U$-$B color.  
HD~37699 also has a very blue U$-$B color excess with
E(U$-$B)=$-$0.$^m$11.  Since these two have values of E(U$-$B) that
are significantly less than zero, we excluded these stars from the
calculation of the median E(U$-$B).  Our median E(U$-$B) is consistent
with our mean E(B$-$V) and a normal reddening law.

The small reddening has a disproportionate impact on main-sequence
fitting on the V, B-V plane because the ZAMS has a slope $\frac{\Delta
V}{\Delta (B-V)}\sim$18 for stars near B5V.  If we were to ignore the
cluster's reddening, we would find a distance that is $\sim$100~pc
smaller than we find when correcting for the observed color excess.

%Since our estimated values of A$_V$ range from 0.06 to
%1.05 (see Table~\ref{true_table}), it is important to correct each
%individual star's data for that star's estimated A$_V$.  Many of the
%stars have values of E(B$-$V) that are less than that star's
%uncertainty on (B$-$V)$_0$, which we estimate by assuming an an
%uncertainty on the spectral type of $\pm$1 subtype.  We do not attempt
%to correct stars for reddening if the star's E(B$-$V) is less than
%twice the uncertainty on (B$-$V)$_0$.

\subsection{The Main Sequence Turn--On }\label{to}

\begin{figure}
%\plotone
\includegraphics[scale=.80]{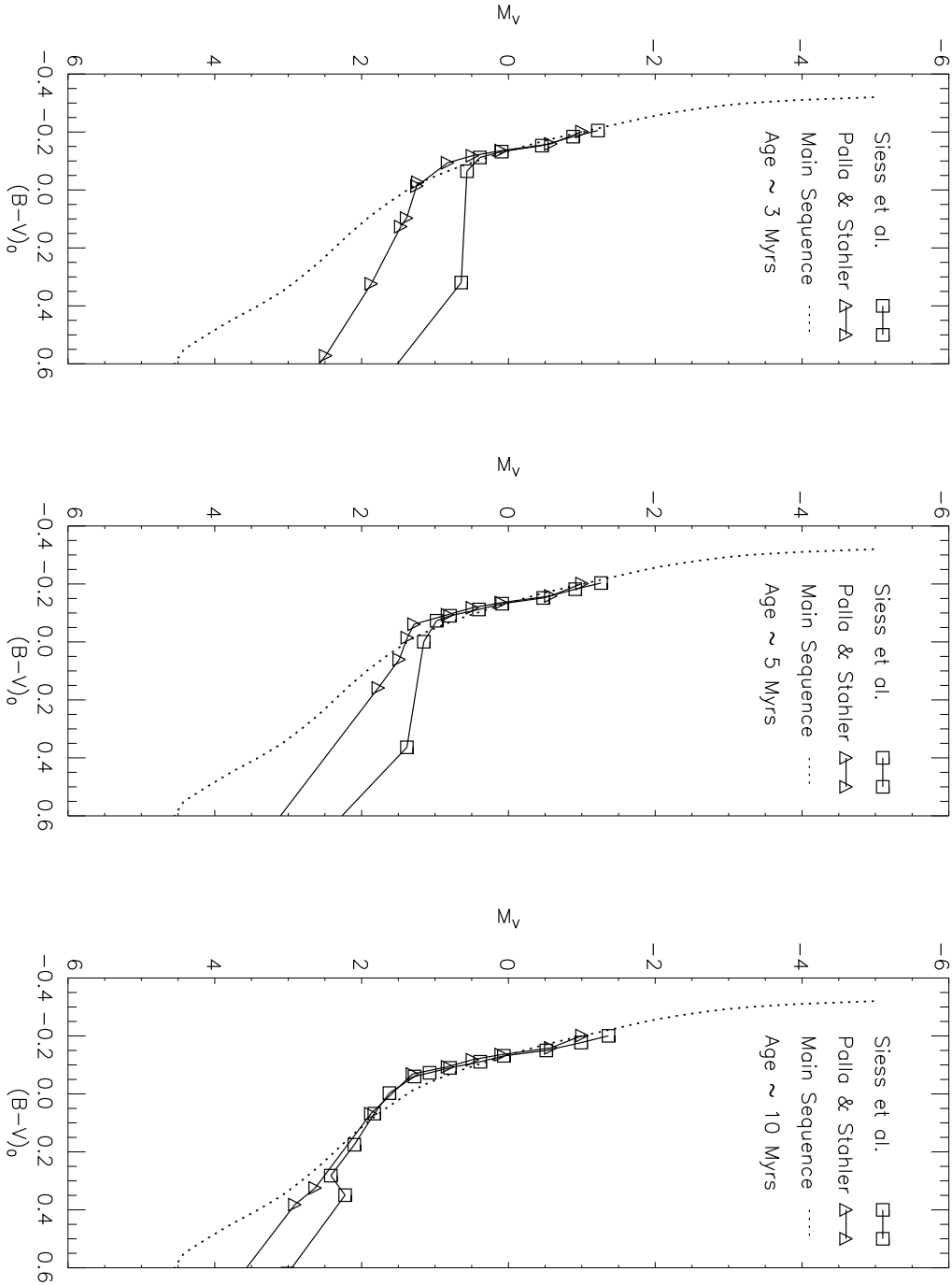}
\caption{The three panels of this figure illustrate the evolution of
the main sequence turn--on from an age of $\sim$3~Myrs (Left) through
an age of $\sim$5~Myrs (Middle) to an age of $\sim$10~Myrs.  At ages
of $\sim$3~Myrs and $\sim$5~Myrs both the \citet{siess00} and the
\citet{ps99} isochrones join the ZAMS \citep{turn76,turn79} near
(B$-$V)$_0$ of 0.0.  The late A--stars are on the ZAMS only for the
$\sim$10~Myr isochrones.
\label{zams}
}
\end{figure}

Since the $\sigma$~Ori cluster is roughly 3~Myrs old, most of the
cluster members have not reached the ZAMS.  Figure~\ref{zams} compares
the ZAMS of \citet{turn76, turn79} to theoretical isochrones from
\citet{siess00} and \citet{ps99}.  These models predict that the main
sequence turn--on is near very late B or very early A spectral type, depending
upon the assumed age and the choice of model.  
%This forces us to use
%only the cluster's B--stars to estimate the distance to the cluster.
Therefore we exclude the A stars in fitting the main sequence, as they are
likely to lie above the ZAMS.

\subsection{The Abundances of $\sigma$~Ori Cluster Members}\label{comp}

%We estimate the distance to the cluster using only the highest mass
%members for two reasons.  At the age of the cluster, only the upper
%main sequence has reached the ZAMS.  Second, it is difficult to
%identify cluster members with late A through early G spectral types
%because of confusion with field stars.

%UBV photometry is available for all of the stars in our sample.  
%This makes it natural to fit the cluster's distance modulus on the V
%vs.\ B$-$V CMD.  
%The M$_V$ of an isochrone depends upon its metallicity.
\citet{cunha4} report that [Fe/H] for the Orion OB association as a
whole is $-$0.16$\pm$0.11.  However, most of the stars in their sample
were from Orion OB1c and 1d.  They also report variations of
the oxygen to iron ratio with position within Orion OB1, possibly due
to self--enrichment by supernovae.  
%**  ??? Ins't this what you discuss below???
%It is not clear what effect, if
%any, a different pattern of abundances would have on the location of
%an isochrone on the CMD.

Metallicity changes the position of stars on the CMD in three ways.
Lower metallicity stars are slightly more luminous (at a fixed mass).
Lower metallicity stars also have smaller radii, and thus higher
temperatures and earlier spectral types at a given mass.  Among
low--mass stars, lower metallicity stars also have less
line--blanketing.  This makes them bluer than higher metallicity stars
of the same spectral type.

%For older clusters the lower main sequence is used for main sequence
%fitting.  The effects of line--blanketing make the measured distance
%modulus more sensitive to the cluster's metallicity.  \citet{pin2004}
%estimate the sensitivity of an isochrone's M$_V$ to metallicity for
%stars with 0.4$\leq$(B$-$V)$_0\leq$1.0.  They find that, on the M$_V$
%vs.\ (B$-$V)$_0$ CMD, $\Delta$M$_V$=1.4$\Delta$[Fe/H].  If we were
%using F, G, and K stars to estimate the distance to the cluster, we
%would need to adjust our isochrone by (+)0.2 magnitudes (0.$^m$2
%fainter) if we assume [Fe/H]=$-$0.16.  This would decrease the best
%fit distance modulus by 0.2~mag.
%%Our sample is mainly
%%(420~pc instead of 450~pc).  
%%(400~pc instead of 440~pc)
%%stars with spectral types of A2V or earlier and we only use the
%%B--stars to estimate the cluster's distance.  

The (U$-$B)$_0$ and (B$-$V)$_0$ colors of early~type stars are much
less sensitive to metallicity than those of late~type stars.  From
Table~1 of \citet{cameron85} it is clear that B stars with only
slightly sub--solar [Fe/H] should have UBV colors that are the same as
those of solar metallicity stars to within 0.$^m$01. This reduces
our sensitivity to [Fe/H].  However, the B stars will still have lower
masses than solar metallicity stars of the same spectral type.  This
will make the ZAMS fainter than the standard ZAMS, but not by as much
as would be the case for later type stars.

\begin{figure}
\plotone{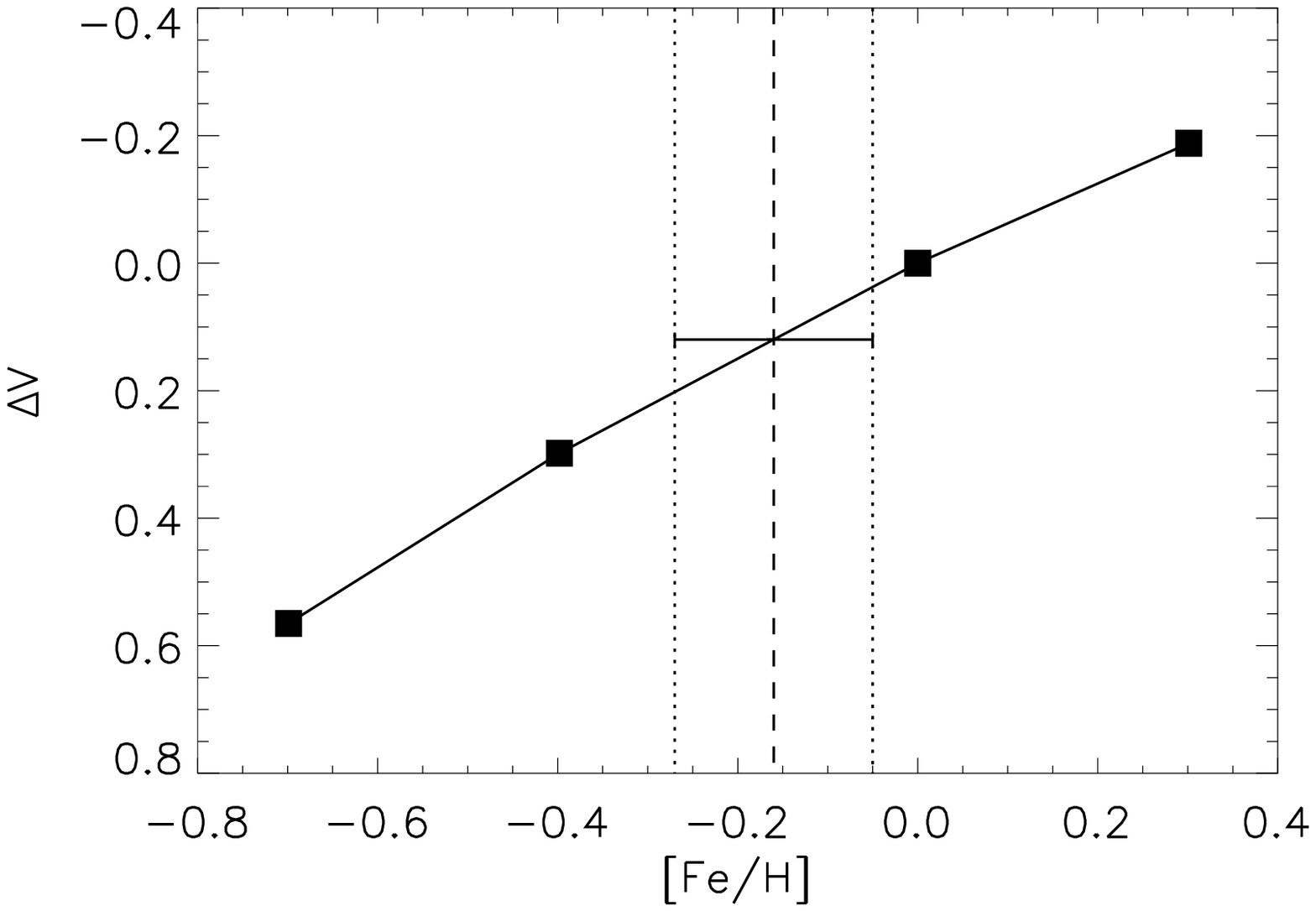}
\caption{The change in the M$_V$ of the 2~Myr isochrones of
\citet{lej2001} at a fixed (B$-$V) color of $-$0.$^m$21 as the
metallicity ([Fe/H]) increases from [Fe/H]=$-$0.7 (z=0.004) to
[Fe/H]=+0.30 (z=0.040).  The dashed line marks the average value of
[Fe/H] for Orion \citep{cunha4}.  The two dotted lines are the
$\pm$1~$\sigma$~values for [Fe/H].  The horizontal bar marks the
predicted $\Delta$V between a solar metallicity isochrone and an
isochrone with [Fe/H]=-0.16.  From this plot we conclude that Orion's
sub--solar metallicity will make the ZAMS fainter by between 0.$^m$04
and 0.$^m$2.
\label{zplot}
}
\end{figure}

We quantify the shift in the ZAMS as a function of [Fe/H] by examining
the change in the M$_V$ of isochrones from the models of
\citet{lej2001} at (B$-$V)$_0$=$-$0.$^m$21 as the metallicity varies
from z=0.040 ([Fe/H]=+0.3) to z=0.004 ([Fe/H]=$-$0.7).
Figure~\ref{zplot} shows that $\Delta$M$_V\simeq-$0.75$\times$[Fe/H].
An isochrone with [Fe/H]=$-$0.16 is 0$^m$.12 fainter than a solar
metallicity isochrone and 0$^m$.15 fainter than an isochrone with the
metallicity of the Pleiades, [Fe/H]=+0.04.  The metallicity of the
Pleiades matters because we use the Pleiades to calibrate the ZAMS, so
our ZAMS is matched to an [Fe/H] of +0.04 (see Section
\ref{which_zams}).  Since the uncertainty on the measured value of
[Fe/H] for Orion is quite large ($-$0.05$<$[Fe/H]$<-$0.3), the
correction to the ZAMS is in the range +0.$^m$07 to 0.$^m$23.  This is
a systematic correction to the distance modulus of
$-$0.$^m$15$\pm$0.$^m$08.

%The largest source of uncertainty on the distance we find from main
%sequence fitting is the uncertainty in the zero age main sequence.

\subsection{Cluster Membership}\label{memb_sec}

%Several of the stars in our sample are much brighter or much
%fainter than the expected locus of cluster members.  We consider
%HD~37564 to be a non--member of the cluster because it lies more than
%1~mag.\ above the 2.5~Myr isochrone.  We consider HD~294273 and
%HD~294279 to be non-members because they lie more than 0.5~mag.\ below
%the 2.5~Myr isochrone.  HD~294272 and HD~37333 both lie
%$\sim$0.7~mag.\ above the 2.5~Myr isochrone.  
%%Given the uncertainties in the exact shape and age of the PMS isochrone, 
%These two stars could be either binary cluster members or foreground
%non--members.

\begin{figure}
\plotone{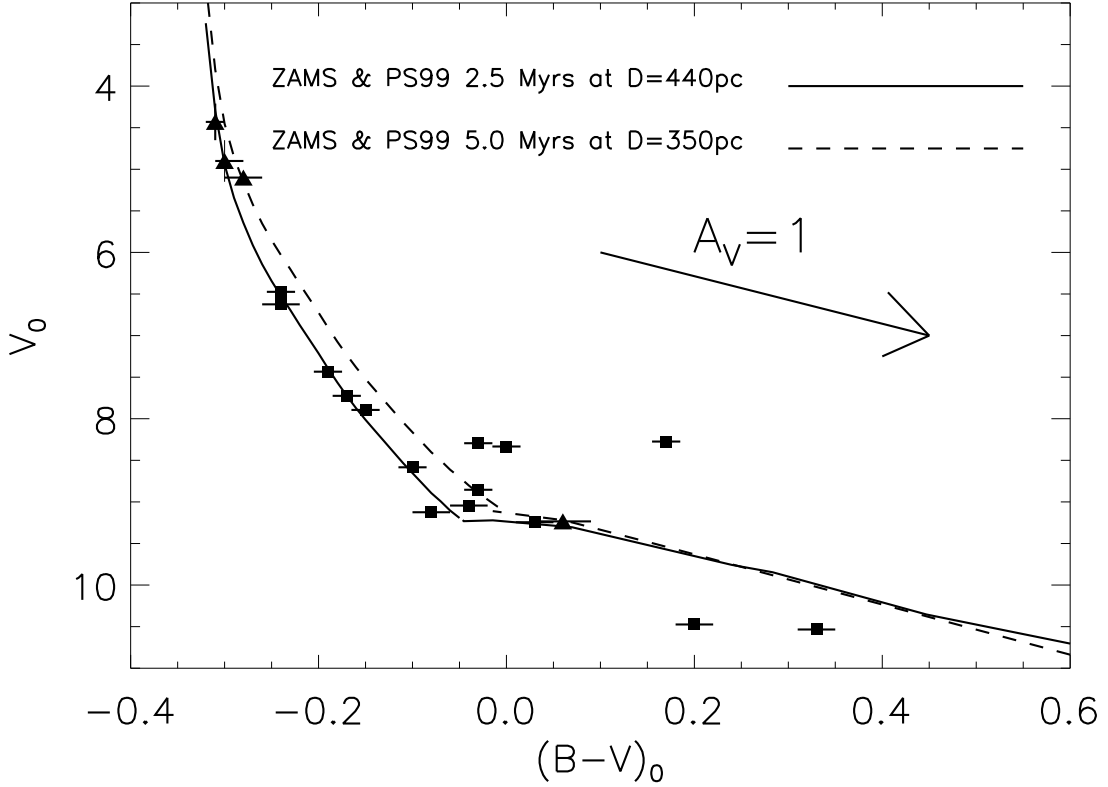}
\caption{This color--magnitude diagram compares the dereddened (B$-$V)
colors and V magnitudes of the early type stars within 30$^{\prime}$
of $\sigma$~Ori~AB to the expected locus of cluster members for
distances of 440~pc and 350~pc.  The solid line is an empirical
solar--metallicity ZAMS \citep{turn76} plus a 2.5~Myr isochrone
\citep{ps99} at a distance of 440~pc.  The 2.5~Myr isochrone matches
the cluster age estimated from the low--mass members \citep{paper1} as
well as the position of $\sigma$~Ori~C.  The dashed line is the ZAMS
at a distance of 350~pc and a 5~Myr isochrone that was chosen to match
the position of $\sigma$~Ori~C.  The triangles mark the positions of
$\sigma$~Ori~Aa, $\sigma$~Ori~Ab, $\sigma$~Ori~B, and $\sigma$~Ori~C.
These four stars were placed on the CMD using values of (B$-$V)$_0$
predicted from their spectral types because no reliable measured
values of (B$-$V) were available.  The squares mark the positions of
the remaining stars from Table~\ref{true_table}.
\label{memb}
}
\end{figure}

Figure~\ref{memb} shows a CMD with the ZAMS and isochrones over
plotted.  Fourteen of the stars included in Figure~\ref{memb} are
consistent with a distance of 420~pc and an [Fe/H] of $-$0.16.  The A
stars HD~37564 and HD~37333 are much brighter than A type cluster
members should be.  This suggests that they may be foreground stars or
binaries.

\subsubsection{HD~37564}

The SIMBAD data base lists this star with a spectral type of A0V.
This spectral type appears to be from the SAO catalog \citep{sao}.
\citet{guetter81} reports a spectral type of A8V, however, the
measured value of B$-$V is slightly bluer than (B$-$V)$_0$ for an A8V
star \citep{kh95}.  If the color is correct, then HD~37564 should have
a spectral type of A7 or earlier if it is reddened. In our spectrum the Balmer
lines match spectral type A5 standards, while the depth of the CaII K line
suggests a slightly later type of about A7.
%*************  spectral discussion added above

HD~37564 is the reddest of the three bright outliers on
figure~\ref{memb}.  At spectral type A7V, it lies roughly 1.4
magnitude above the 2.5~Myr isochrone.  HD~37564 could be a foreground
field star at a distance of $\sim$150~pc.  If the spectral type is
later than A7V, then the mismatch between the measured V magnitude and
the expected V magnitude for a distance of 440~pc is even larger.
This is consistent with previous studies that concluded that HD~37564
is probably not a member of Orion OB1 \citep{brown94}.

%If HD~37564 is in fact a cluster member, it would have to be both a
%B9V star and an equal mass binary.  Since A stars are relatively
%common, we think it is very likely that HD~37564 is a foreground field
%star.  

\subsubsection{HD~37333}

HD~37333, the other obvious outlier, is 0.75 magnitudes
brighter than expected for a dereddened A0V star at a distance of
440~pc.  Since 0.75 magnitudes is the exact magnitude difference which
an equal mass binary would have, it is quite plausible that HD~37333
is a cluster member that is an equal mass binary.

If HD~37333 is not an equal mass (or nearly equal mass) binary, then
it is too bright to be a member of the cluster.  %In that case its 
If it is a main sequence field star, its likely distance modulus would be
$\sim$310~pc.  A distance of 310~pc would be consistent with
membership in the Orion OB1a group, but an A0V ZAMS star would be
more than 0.5 mag.\ fainter than an A0V main sequence field star.

\subsubsection{HD~294272}

HD~294272 is a member of a multiple system, ADS~4240.  It is not clear
if ADS~4240A (HD~294271), ADS~4240B (HD~294272), and ADS~4240C
(BD~$-$02~1323C) are a bound system or a chance alignment.  The small
separation between HD~294272 and BD~$-$02~1323C
(8.5$^{{\prime}{\prime}}$) suggests that at least these two stars are
a physical system.  However, HD~294272 has been classified as a
B9.5III star \citep{guetter81} and is 0.$^m$29 brighter than
BD~$-$02~1323C which is a B8V star.  HD~294272 lies $\sim$0.$^m$8
above the 2.5~Myr isochrone \citep{ps99}.  The location of
BD~$-$02~1323C on the CMD is consistent with the best fit distance to
the cluster.  If HD~294272 and BD~$-$02~1323C are in fact a bound
pair, then HD~294272 is too bright to be a main~sequence star.  One
possibility is that HD~294272 is a binary with a PMS companion.  This
would make the unresolved binary brighter and redder.  Since the PMS
companion would be brighter than a main~sequence star of the same
color, the unresolved system could be consistent with cluster
membership.  An alternative is that HD~294272 is not a cluster member,
although this makes the small apparent separation unlikely.  %In this
%case, like HD~37333, its position on the CMD would be consistent with
%membership in the Orion OB1a group.

%The other possiblity is that HD~294272 is an
%unresolved binary that has been assigned the wrong luminosity class.

\subsubsection{HD~294279}\label{HD294297}

%******************** revised paragraph
As stated above, we find that HD~294279 has a spectral type of
F0-F3. This is consistent with the observed B$-$V color of 0.39 and
modest reddening. An A$_V$ of $\sim$0 places the star $\sim$1
magnitude above the main sequence and ~1~magnitude below the 2.5~Myr
isochrone at the best fit distance to the cluster.  This strongly
suggests that HD~294279 is a foreground field star.

\citet{cab_thesis} report a spectral type of F3 to F5 and the
detection of Li in the spectrum.  If the spectral type were as late as
F5, then HD~294279 would lie near the isochrone, but its observed
B$-$V is 0.39 which is much too blue for an F5 star with the clusters
measured E(B$-$V).  Such a star should have B$-$V=0.50.  The detection
of Li in the spectrum suggests that HD~294279 may be a member of
Orion~OB1a.  HD~294279 lies closer to an 11~Myr old isochrone at
330~pc than it does to the $\sigma$~Ori isochrone.

%This would imply an A$_V$ of 1.02.  This would make the dereddened
%position on Figure~\ref{true_cmd} consistent with the best fit
%distance for a cluster member.

\subsubsection{HD~294273}

%******************** revised paragraph
With a revised spectral type of A7, 
%HD~294273 has an HD spectral type of A3.  \citet{walter07} find that
%HD~294273 is an A7 to A9 star.  Assuming a (B$-$V)$_0$ color that is
%appropriate for a late A--star (see Table~\ref{true_table}), 
HD~294273
has an A$_V$ which is the same as that of cluster members.  It
also lies on the ZAMS for a distance of 440~pc.  However, for it to be
a cluster member it would need to be $\sim$5~Myrs older than the age
of the cluster.  Since there is no evidence for such a large age
spread, we think that it is far more likely that HD~294273 is a field
star.  It is probably slightly more distant than the $\sigma$~Ori
cluster because stars on the main sequence are brighter than stars of
the same spectral type on the ZAMS.

\subsection{The Cluster's Distance}\label{dist_sec}

Fourteen of the 19 O, B, and A stars in our sample lie close to the
2.5~Myr isochrone on Figure~\ref{memb}.  Of these, only 10 of the O
and B--stars lie on the main~sequence.  Three of these stars,
$\sigma$~Ori Aa, Ab, and B, do not have directly measured colors.  We
have used the observed spectral types of $\sigma$~Ori~Aa and
$\sigma$~Ori~B to estimate their (U$-$B)$_0$ and (B$-$V)$_0$ colors,
but there are no measurements of the spectral type or colors for
$\sigma$~Ori~Ab.  This makes $\sigma$~Ori~Ab unusable, leaving us with
9 usable main sequence cluster members.  The stars which we have used
to determine the cluster's distance are marked with an asterix in
column 11 of Table~\ref{true_table}.
%Therefore, we will not use it to estimate the cluster's
%distance.

We estimate the best fit distance to the $\sigma$~Ori~cluster by
calculating ${\chi}^2_{\nu}$ for the colors and magnitudes from
Table~\ref{true_table} compared to two versions of the ZAMS shifted
to a series of distances.  We estimated the uncertainty on the
best fit distances as the distances at which we found that
${\chi}^2={\chi}^2_{best}$+1.

%The result is shown in Figure~\ref{chiplot}.
%In Figure~\ref{chiplot}, 
%We used the colors and magnitudes from all of the stars in
%Table~\ref{true_table}, except probable non-member HD~37564 and
%possible binary HD~37333.  The distance which yields the smallest
%${\chi}^2_{\nu}$, 0.43 for $\nu$=1 degree of freedom, is
%442$^{+22}_{-19}$~pc. 
%If we assume uncertainties of 0.5 subtypes, we get the same
%best distance with an uncertainty of $\sim$16~pc and a
%${\chi}^2_{\nu}$ of 0.97.

\subsubsection{Choice of ZAMS}\label{which_zams}

%Figure~\ref{ms3} illustrates the difference in the shape of a 2.5~Myr
%isochrone from \citet{lej2001} and 
%We examined two empirical ZAMS \citep{turn76,turn79,sk82} and an
%isochrone from \citet{lej2001} in order to select the best
%representation of the $\sigma$~Ori~cluster's main sequence.  The two
%empirical ZAMS have very similar shapes and are in good agreement with
%each other.  Both ZAMS closely follow the cluster's main sequence.
%The model isochrone has a shape which is basically similar to the
%empirical ZAMS, but is notably different for values of
%(B$-$V)$_0<$0.1.  The better fit of the shape of the empirical ZAMS to
%the observed V magnitudes and (B$-$V)$_0$ colors of the cluster
%members leads us to place greater trust in the empirical ZAMS.  We
%choose to use only the two empirical ZAMS to estimate the best fit
%distance to the cluster.  

We examined two empirical ZAMS \citep{turn76,turn79,sk82} in order to
select the best representation of the $\sigma$~Ori~cluster's main
sequence.  Both ZAMS closely follow the cluster's main sequence.  

We use both versions of the ZAMS because although Turner's ZAMS
follows the locus of cluster members on the CMD nearly perfectly, it
does not include U band photometry.  The ZAMS from Schmidt--Kaler does
not trace the locus of our data quite as well, but includes U band
photometry.
%, but does have U band photometry.
%*** paragrapg moved from sec 2.6.3
The U band ZAMS values are valuable because the slope of the ZAMS is
$\sim$6 on the (U$-$B)$_0$ vs.\ V$_0$ CMD and $\sim$4.5 on the
(U$-$V)$_0$ vs.\ V$_0$ CMD.  These slopes are much shallower than the
slope of $\sim$18 on the (B$-$V)$_0$ vs.\ V$_0$ CMD.  This makes the
best fit distance less sensitive to small errors in the reddening
correction.

%For comparison, we have also plotted the cluster's
%main sequence members.  
%The shape of both empirical ZAMS are a much better match to the locus
%of points traced out by the cluster members than is the model
%isochrone.  
%Due to the
%noticeably differences between the shape of the model isochrone and
%both empirical ZAMS, 

\subsubsection{The Empirical ZAMS of Turner}

%We first choose to compare the data with an empirical ZAMS
%\citep{turn76,turn79}.  
We shifted the ZAMS of \citet{turn76,turn79} to match the Pleiades
cluster which has in turn been matched to the Hyades cluster
\citep{an2007}.  The Hyades cluster has [Fe/H]=+0.14$\pm$0.05 and a
distance modulus of 3.33$\pm$0.01 \citep{perryman98}.  The Pleiades
has a nearly solar [Fe/H] of $\sim$+0.04$\pm$0.03 \citep{an2007}.

The left panel of Figure~\ref{true_cmd} shows the 10 members of the
cluster which have reached the ZAMS.  The lines show the ZAMS of
Turner shifted to distances ranging from 350~pc to 464~pc.

We estimated the most probable distance to the cluster by calculating
${\chi}^2_{\nu}$ for the ZAMS shifted to 1000 candidate distances
between 280~pc and 530~pc.  For each candidate distance we calculated
${\chi}^2_{\nu}$ from the separation between the empirical ZAMS and
the (B$-$V)$_0$ colors of the nine main~sequence cluster members for
which we have directly measured B$-$V colors, or spectral types.  We
used the (B$-$V)$_0$ colors as the dependent variable for the
${\chi}^2_{\nu}$ calculation because the uncertainties in the colors
dominate the uncertainties on the estimated distance.  Our best fit
distance is 442$\pm$20~pc.

The small value, 0.5, of our best ${\chi}^2_{\nu}$ suggests that we
have over--estimated the uncertainties on the (B$-$V)$_0$ colors.  

%An over--estimate of the uncertainties is also suggested by the fact
%that most of the points on Figure~\ref{true_cmd} lie within half a
%$\sigma$ of the ZAMS.

%\citet{turn76,turn79} do not include U band data for the ZAMS.  As a
%result, it is not possible to use Turner's ZAMS to estimate the
%cluster's distance using a V, U$-$B or V, U$-$V cmd.  Including U band
%data is useful because the slope of the ZAMS is $\sim$5 on the V,
%U$-$B CMD for spectral types near B5, and $\sim$4 on the V, U$-$V CMD.
%Fortunately, the ZAMS of \citet{sk82} does include U band data.

\subsubsection{The Empirical ZAMS of Schmidt--Kaler}

Using the ZAMS of \citet{sk82}, also shifted to match the Pleiades, on
the (B$-$V)$_0$ vs.\ V$_0$ CMD, we find a best fit distance of
462$^{+14}_{-35}$~pc.  This is 1~$\sigma$ larger than the best fit
distance using the \citet{turn76} ZAMS.

%\citet{turn76} provides values only for (B$-$V)$_0$ and M$_V$.
%\citet{sk82} provides (U$-$B)$_0$, (B$-$V)$_0$, and M$_V$.  This is
%very useful because the slope of the ZAMS is $\sim$6 on the
%(U$-$B)$_0$ vs.\ V$_0$ CMD and $\sim$4.5 on the (U$-$V)$_0$ vs.\ V$_0$
%CMD.  These slopes are much shallower than the slope of $\sim$18 on
%the (B$-$V)$_0$ vs.\ V$_0$ CMD.  On CMDs where the ZAMS has a
%shallower slope, the best fit distance modulus has much less
%sensitivity to small errors in the reddening correction.

The right panel of Figure~\ref{true_cmd} compares the data for the
main~sequence cluster members to the ZAMS on the (U$-$B)$_0$ vs.\
V$_0$ CMD.  The best fit distance is 488$^{+18}_{-20}$~pc.
%Since the mean measured value E(U$-$B) is much less precise than the measured mean E(B$-$V),
The right panel of Figure~\ref{true_cmd} plots (U$-$B)$_0$ colors
corrected using the expected value of E(U$-$B)=0.04.  The V data have
been corrected for reddening using the A$_V$ derived from the mean
E(B$-$V) in Section~\ref{red}.  If the (U$-$B) colors were corrected
by the mean value of E(U$-$B)=0.02, then the derived distance would be
458$^{+15}_{-17}$~pc.

%Since this would not be consistent with the
%observed relationship between spectral type and (U$-$B)$_0$ and the
%ZAMS \citep{sk82}, we assume that the measured E(U$-$B) is zero due to
%either a zero--point error in the U band photometry, or because the
%B--stars have a U band excess.  By using the observed values for
%(U$-$B), we are in effect using a (U$-$B)$_0$ scale set by the
%observed spectral types.

%Figure~\ref{uvv_cmd} compares the data for the main~sequence cluster
%members to the ZAMS on the (U$-$V)$_0$ vs.\ V$_0$ CMD.  
We also estimated the cluster's distance on the (U$-$V)$_0$ vs.\ V$_0$
CMD. The best fit distance is 492$^{+18}_{-31}$~pc using a normal
reddening law and the observed E(B$-$V) of 0.$^m$06$\pm$0.005, or
457$^{+28}_{-16}$~pc using the measured value of
E(U$-$V)=0.$^m$08$\pm$0.03.  The distances estimated using the U band
photometry combined with a standard reddening law are 2$\sigma$ larger
than the distances estimated from colors corrected by the mean
measured E(U$-$B) or E(U$-$V).  They are also 2$\sigma$ larger than
the distance estimated from the (B$-$V) colors.  This may indicate
that the cluster has a non--standard reddening law with
E(U$-$B)=0.36E(B$-$V).

An alternative is to use the reddening corrected 
(U$-$B)$_0$ and (U$-$V)$_0$ colors that correspond to the observed
spectral types.  Doing that, we find distances of 452$\pm$14~pc and
447$^{+13}_{-7}$~pc respectively.  Since there is much more scatter in
the observed U$-$B colors (which we used to get the U$-$V colors) than
in the observed B$-$V colors, using the spectroscopic colors to
estimate the cluster's distance may be more accurate.

Table~\ref{dist_table} collects all of the distances we estimated
using different color indices, reddenings, ZAMS, and metallicities.

\subsubsection{The Best Distance Estimate}

%The results we found
%using the \citet{sk82} ZAMS compared to the three UBV CMDs and the
%\citet{turn76} ZAMS we examined should not 

It is not straight--forward to combine the best--fit distance
estimates into a single best--fit distance.  The various results we
obtained using the \citet{sk82}) and the \citet{turn76} ZAMS should
not simply be averaged because the differences between them are
systematic, not random.  Given the difficulties correcting for the
reddening on the V, U$-$B and V, U$-$V CMDs, we give more weight to
the distance estimates we found using the spectroscopic colors.  These
distance estimates, combined with the distance estimates from the V,
B$-$V CMD, indicate a best--fit distance of 444$\pm$20~pc assuming
solar metallicity.  Correcting for the lower metallicity of the
cluster brings the distance estimate down to somewhere between 400~pc
and 440~pc.  At the most likely metallicity, the best--fit cluster
distance is 420$\pm$20~(random)$\pm$25~(systematic)~pc or
420$\pm$30~pc.  As knowledge of the cluster's metallicity is refined,
the best estimate of the distance modulus will shift and the
systematic component of the uncertainty will be reduced.  It is
important to note that for the purposes of comparing the cluster to
solar metallicity isochrones, the isochrone should be shifted to be
$\sim$0$^m$.12 fainter to compensate for the lower luminosity of a
sub--solar metallicity isochrone.  For low--mass stars
line--blanketing will shift the isochrone by an additional amount that
will depend upon the color index used on the CMD.

The final line of Table~\ref{dist_table} lists the best ``average''
distance estimates for five assumed values for the cluster's
metallicity.  The assumed metallicities range from +0.04 to $-$0.27.
The uncertainties include only the random component of the
uncertainty.

\section{Discussion}

An improved distance estimate for the $\sigma$~Ori cluster will impact
the conclusions of previous research by varying degrees, depending
upon what distance the authors assumed.

%The choice of ZAMS could be an important source of uncertainty in the distance to the cluster.  

\subsection{Implications for the Age of the Cluster}

The position of $\sigma$~Ori~C on figure~\ref{memb} suggests a cluster
age of $\sim$2.5~Myrs.  This is consistent with the age of 2--3~Myrs
estimated from the low--mass stars by \citet{paper1}.  \citet{paper1}
did not correct for the small reddening, or the sub--solar metallicity
of the cluster.  An A$_V$ of 0.19 would make the low--mass stars a bit
brighter and therefore slightly younger.  A sub--solar metallicity
should make the low--mass stars bluer at a given spectral type.  This
would make the cluster's locus on the CMD fainter (thus a bit older).
The net effect on the estimated age is likely to be small, but without
isochrones matched to the cluster's metallicity, the net effect is
difficult to estimate.

\subsection{Adjustments to the Cluster's X--Ray Luminosity Distribution Function}

\citet{xmm2} compared the X--Ray Luminosity Distribution Function
(XLDF) of candidate low--mass members of the $\sigma$~Ori cluster to
the XLDFs of $\rho$~Oph, the Orion nebula cluster (ONC), and the Cha~I
star forming regions.  They found that the median 0.1--4~keV luminosity
of K--type candidate members of the $\sigma$~Ori cluster is a factor
of 3 to 5 fainter than that for the ONC or Cha~I, but comparable to
that of $\rho$~Oph.  They also found that the median X--ray luminosity
for $\sigma$~Ori cluster M--type candidate members is significantly
lower than those of all three star forming regions.  \citet{xmm2} note
that the discrepancy between the observed $\sigma$~Ori XLDF and the
ONC or Cha~I XLDFs may be due in part to contamination of the
membership list by field stars.  They report that their sample
includes 45 candidate $\sigma$~Ori cluster member that were selected
from just photometric data and that were not detected with XMM.  The
discrepancy between the XLDFs of the ONC and Cha~I and the XLDF of the
$\sigma$~Ori cluster is significantly reduced if most or all of these
45 candidate cluster members are assumed to be non-members.

\citet{xmm2} assumed the Hipparcos distance of 352~pc for $\sigma$~Ori.
Also, they used an XLDF for the ONC that assumed a distance of
470~pc \citep{flac2003}.  Recent work shows that the ONC is at a
distance of 414$\pm$7~pc \citep{oncd}.  Using our best fit distance of
420~pc and the new ONC distance, the ratio of the median X--ray
luminosities would be larger by nearly a factor of 2.  This, combined
with the high probability that many of the 45 photometric candidate
members that were not detected with XMM are non--members, may account
for the apparent faintness of $\sigma$~Ori cluster members relative to
the ONC.

\subsection{$\sigma$~Ori~B}

Since $\sigma$~Ori~B is observed to orbit $\sigma$~Ori~A
\citep{heintz1974, heintz1997, frost1904}, it is at the same distance as
$\sigma$~Ori~Aa and Ab.  Yet, $\sigma$~Ori~B is consistently $\sim$0.$^m$6
brighter than the ZAMS which fits all of the other high--mass
cluster members.  This suggests that $\sigma$~Ori~B is a close
binary just as $\sigma$~Ori~A is.  Alternatively, our assumed colors
could be slightly wrong.  If $\sigma$~Ori~B is a bit hotter and bluer
than expected for a B0.5V star, it would be more consistent with the
best fit ZAMS.  A sufficiently blue value of (B$-$V)$_0$ would be
within 1$\sigma$ of the standard color of a B0.5V.  However,
sufficiently blue (U$-$B)$_0$ or (U$-$V)$_0$ colors would be difficult
to reconcile with the observed B0.5V spectral type.  We feel that a
close binary companion is the more likely explaination.

\section{Conclusions}

\begin{figure}
\plotone{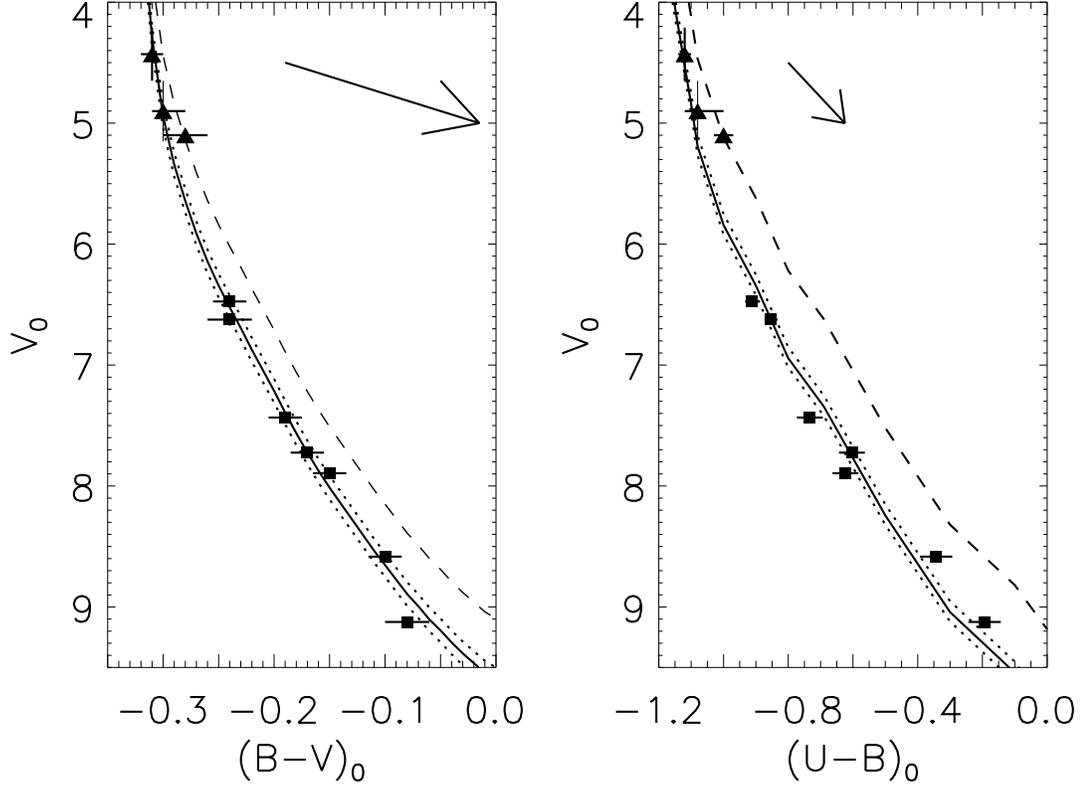}%sori_msfit_v5.eps}
\caption{{\bf Left}: This color--magnitude diagram compares the
corrected and derived values of (B$-$V)$_0$ and V magnitude for likely
cluster members (see Table~\ref{true_table}) along with the
solar--metallicity zero age main sequence of \citet{turn76} shifted to
the best fit distance (solid line), the best fit $\pm$1$\sigma$
distances (dotted lines), and the Hipparcos distance (dashed line).
The triangles mark the colors and magnitudes derived for
$\sigma$~Ori~Aa, $\sigma$~Ori~Ab, and $\sigma$~Ori~B.  The color of
$\sigma$~Ori~Ab was derived by assuming that it lies on the ZAMS at
the same distance as $\sigma$~Ori~Aa, so $\sigma$~Ori~Ab was not used
to derive the best fit distance.  The filled squares show the
dereddened colors and magnitudes for the remaining main~sequence
B--stars identified as cluster members in Table~\ref{true_table}.  The
arrow shows a reddening vector for A$_V$=0.$^m$5.  {\bf Right}: This
color--magnitude diagram compares the corrected and derived values of
(U$-$B)$_0$ and V magnitude for likely cluster members (see
Table~\ref{true_table}) along with the zero age main sequence of
\citet{sk82} shifted to the best fit distance (solid line), the best
fit $\pm$1$\sigma$ distances (dotted lines), and the Hipparcos
distance (dashed line).  The plot symbols are the same as on the right
panel.  The arrow shows a reddening vector for A$_V$=0.$^m$5.
\label{true_cmd}
}
\end{figure}

From Figure~\ref{true_cmd} it is clear that the $\sigma$~Ori cluster
must be more distant than the nominal 350~pc Hipparcos distance for
$\sigma$~Ori.  We estimate a distance of 420$\pm$30~pc for the
cluster, assuming an [Fe/H] %[$\frac{Fe}{H}$] 
of $-$0.16, or 444~pc assuming solar metallicity.  This is consistent
with, but significantly more precise (7\%) than the Hipparcos distance
(30\%).  Our distance estimate is consistent with the estimated
distance to Orion OB1b.  Most of the older age estimates for the
cluster assumed a distance of 350~pc.  Our more tightly constrained
distance shows that the cluster age must be closer to the young end of
the range of the estimated ages.  This places the age of the cluster
in the range of 2--3~Myrs.

%If the cluster has the same, sub--solar metallicity that other regions
%of the Orion OB1 association have, then the best--fit cluster distance
%is 420~pc.  Even if this is the case, 
%research using solar metallicity isochrones
%should use the solar metallicity distance of 444~pc to derive
%ages because the larger distance will partially compensate for the use of solar
%metallicity in the models.

Sixteen of the 19 stars in our sample are probable members of the
$\sigma$~Ori cluster. HD~37564 is too bright to be a cluster member.
%consistent with the other stars in our sample unless it is an equal
%mass binary with a spectral type of B9V.  This seems a bit unlikely
%given its observed spectral type.  
HD~294279 is too faint to be a cluster member. 
%given its spectral type of F3.  We suspect that HD~37333 may be an equal mass binary.  
If HD~37333 is in fact an equal mass binary, it is likely to be a
cluster member.  %Otherwise, it may be part of Orion~OB1a.

The existence of a tight main sequence among the O, B, and A
stars of the $\sigma$~Ori cluster suggests that any other clusters
within the Orion OB1a and Orion OB1b groups (such as the 25~Ori
cluster \citep{briceno05}) should also have tight main sequences.  

\acknowledgments

This research has made use of the SIMBAD database, operated at CDS,
Strasbourg, France.  Stony Brook University's participation in the
SMARTS consortium is made possible by generous contributions by the
Vice President for Research, the Provost, and the Dean of Arts and
Sciences.

WHS was supported in part by the NAI under Cooperative Agreement
No. CAN-02-OSS-02 issued through the Office of Space Science.  The NSO
and the NOAO are operated by AURA for the National Science Foundation.

SJW was supported by NASA contract NAS8-03060.

\clearpage

%% the file, uncomment the \plotone, \plottwo, and
%% If you need a layout that cannot be achieved with \plotone or
%% \plottwo, you can invoke the graphicx package directly with the
%% \includegraphics command or use \plotfiddle. For more information,
%% please see the tutorial on "Using Electronic Art with AASTeX" in the
%% documentation section at the AASTeX Web site,
%% http://www.journals.uchicago.edu/AAS/AASTeX.
%%
%% The examples below also include sample markup for submission of
%% supplemental electronic materials. As always, be sure to check
%% the instructions to authors for the journal you are submitting to
%% for specific submissions guidelines as they vary from

%% This example uses \plotone to include an EPS file scaled to
%% 80% of its natural size with \epsscale. Its caption
%% has been written to indicate that additional figure parts will be
%% available in the electronic journal.

%\begin{figure}
%\epsscale{.80}
%\plotone{sori_msfit_raw2.eps}
%\caption{This Color-magnitude diagram compares the observed B$-$V
%color and V magnitudes for the stars in Table~\ref{raw_table} along
%with the main sequence \citep{bm98} shifted to various distances.
%\label{raw_cmd}}
%\end{figure}

\clearpage

%% Here we use \plottwo to present two versions of the same figure,
%% one in black and white for print the other in RGB color
%% for online presentation. Note that the caption indicates
%% that a color version of the figure will be available online.
%%

%\begin{figure}
%\plotone{dist_chi.eps}
%\caption{This plot of ${\chi}^2_{\nu}$ vs distance shows ${\chi}^2_{\nu}$ 
%calculated by comparing the postions on the CMD of all the stars from 
%Table~\ref{true_table}, except for probably non--member HD~37564, with 
%the main sequence \citep{bm98}.  The minimum ${\chi}^2_{\nu}$, 0.51, is 
%for a distance of 440$^{+27}_{-20}$~pc.  The dashed horizontal line marks 
%the values of ${\chi}^2_{\nu}$ that correspond to $\pm$1$\sigma$ 
%(${\chi}^2$=${\chi}^2_{min}$+1) \citep{}.  The vertical dotted lines 
%mark the $\pm$1$\sigma$ values for the cluster's distance, 420~pc and 467~pc 
%respectively.
%\label{chiplot}}
%\end{figure}

%\begin{figure}
%\plotone{bvv_lejeune_turner_comp.ps}%comp_main_seq.eps}
%\caption{This color--magnitude diagram illustrates the differences
%between three estimates of the upper main sequence.  The dotted line
%\citep{sk82} and the solid line \citep{turn76,turn79} are zero age
%main sequences estimated from observations of young clusters.  The
%distances to these clusters were ultimately derived from measured
%distance to the Hyades and are consistent with a distance modulus of
%3.$^m$33 for the Hyades \citep{perryman98}.  There is very good
%agreement between these two estimates of the main sequence for early
%type stars.  For late type stars (B-V$>$0.$^m$4), these two main
%sequences are nearly identical.  The dashed line is a 2.5~Myr isochrone
%from the models of \citet{lej2001} (starting on the
%ZAMS).  This isochrone matches Turner's ZAMS reasonably well for late
%type stars near (B$-$V)$_0\sim$0.$^m$4, but is up to 0.$^m$1 bluer (or
%0.$^m$4 fainter) for B--stars.
%\label{ms3}
%}
%\end{figure}

%ms_turn_on.v2.ps
%comp_main_seq.eps

\clearpage

\clearpage

\begin{deluxetable}{lcccccccccll}
\tabletypesize{\scriptsize}
\rotate
\tablecaption{Adopted Values\label{true_table}}
\tablewidth{0pt}
\tablehead{
\colhead{ID} & \colhead{U$-$B} & \colhead{(U$-$B)$_0$}\tablenotemark{b} & \colhead{B$-$V} & \colhead{(B$-$V)$_0$}\tablenotemark{b} & \colhead{Err} & \colhead{V}\tablenotemark{1} &
\colhead{Sp.\ Type}\tablenotemark{2} & \colhead{D($^{\prime}$)} & \colhead{A$_V$} & \colhead{Member} & \colhead{Notes          }
}
\startdata
$\sigma$~Ori~Aa  & --      & $-$1.12\tablenotemark{h} & --    & $-$0.31 & 0.01 &   4.4\tablenotemark{3}  &  O9V\tablenotemark{a}   &  0.00 &  0.19 & MS* & Calculated \\
$\sigma$~Ori~Ab  & --      & $-$1.08\tablenotemark{h} & --    & $-$0.30 & 0.01 &   4.9\tablenotemark{3}  &  [B0V?]\tablenotemark{8}   &  0.00 &  0.19 & MS & Calculated \\
$\sigma$~Ori~B   & --      & $-$1.00 & --      & $-$0.28 & 0.02 &   5.16\tablenotemark{3}  &  B0.5V\tablenotemark{a} &  0.00 &  0.19 & MS* & Calculated \\
$\sigma$~Ori~C   & --      & $+$0.06 & --      & $+$0.06 & 0.03 &   9.42 &  A2V   &  0.20 &   --  & PMS & V from \citet{paper3}  \\
$\sigma$~Ori~D   & $-$0.81 & $-$0.84 & $-$0.18 & $-$0.24 & 0.02 &   6.81 &  B2V   &  0.22 &  0.19 & MS* & \\
$\sigma$~Ori~E   & $-$0.87 & $-$0.84 & $-$0.18 & $-$0.24 & 0.014 &  6.66 &  B2Vp  &  0.69 &  0.19 & MS* & Peculiar, Variable \\
HD~294272        & $-$0.05 & $-$0.10 & $+$0.03 & $-$0.04 & 0.015 &  8.48 &  B9.5III\tablenotemark{9}  &  3.12 &  0.22 & PMS? & ADS4240B \\ %\tablenotemark{3} \\
BD~$-$02~1323C   & $-$0.30 & $-$0.37 & $-$0.04 & $-$0.11 & 0.02 &   8.77 &  B8V\tablenotemark{g} & 3.25 & 0.22 & MS* &  \\
HD~294271        & $-$0.56 & $-$0.58 & $-$0.11 & $-$0.17 & 0.01 &   7.91 &  B5V\tablenotemark{c}   &  3.47 &  0.19 & MS* & ADS4240A;  ADS4240B is 68$^{{\prime}{\prime}}$ away. \\
HD~37525         & $-$0.58 & $-$0.58 & $-$0.09 & $-$0.17 & 0.01 &   8.08 &  B5V\tablenotemark{5} &  5.11 &  0.25 & MS*  & May be B5III\tablenotemark{5}; has a faint 0.$^{{\prime}{\prime}}$45 companion\tablenotemark{4} \\
HD~294273        & +0.07   & $+$0.07 & $+$0.26 & $+$0.2  & 0.03 &  10.66 &  A7--9\tablenotemark{6}    &  8.68 &  0.19 & No & HDE Spectral Type:  A3 \\
HD~37564         & +0.15   & $+$0.10 & $+$0.23 & $+$0.17 & 0.03 &   8.46 &  A5/7  &  8.74 &  0.19 & No  & $>$1~mag. brighter than isochrone. \\
HD~37633         & $-$0.36\tablenotemark{7} & $-$0.20 & $+$0.03 & $-$0.06 & 0.04 &   9.04 &  B9   & 16.00 &  0.28 & PMS & Variable (V1147~Ori) \\ %; May have a mid--Bp spectral type\tablenotemark{6} \\
HD~37333         & $-$0.06 & $-$0.07 & $+$0.06 & $+$0.00 & 0.02 &   8.52 &  A0V   & 18.60 &  0.19 & PMS? & Binary or Non--Member\\
HD~294279        & +0.03   & $+$0.01 & $+$0.39 & $+$0.37 & 0.03 &  10.72 &  F3\tablenotemark{6} & 19.34 &  0.06 & OB1a? & See Section~\ref{HD294297} \\
HD~294275        & +0.01   & $+$0.07 & $+$0.09 & $+$0.03 & 0.03 &   9.43 &  A1V\tablenotemark{g}   & 20.45 &  0.19 &  PMS & \\
HD~37545         & $-$0.15 & $-$0.20 & $-$0.02 & $-$0.06 & 0.02 &   9.31 &  B9V   & 21.46 &  0.12 & MS* & \\
HD~37686         & $-$0.09 & $-$0.10 & $+$0.02 & $-$0.04 & 0.015 &  9.23 &  B9.5V & 22.64 &  0.19 & (P?)MS & \\
HD~37699         & $-$0.69 & $-$0.58 & $-$0.13 & $-$0.17 & 0.01 &   7.62 &  B5V   & 25.79 &  0.12 & MS?* & Radial velocity may be inconsistent with membership\tablenotemark{4} \\
\enddata
\tablenotetext{1}{These values are not corrected for extinction, except for $\sigma$~Ori~Aa, Ab, and B. }
\tablenotetext{2}{Spectral types are from \citet{mich5} except where otherwise noted. }
\tablenotetext{3}{These are reddening corrected magnitudes.}
%\tablenotetext{3}{ADS4240C is $\sim$2 magnitudes fainter and separated by $\sim$8.5$^{{\prime}{\prime}}$}
\tablenotetext{4}{\citep{cab2007}}
%\tablenotetext{5}{\citep{mich5}}
\tablenotetext{5}{\citet{mich5} list a spectral type of B5III, but point out that the differences between class V and class III are subtle in B stars.  We have adopted the spectral classification from \citet{schild71}.}
\tablenotetext{6}{Spectral types from our SMARTS 1.5m observations.  See Section~\ref{analysis}}
\tablenotetext{7}{This star seems to have a U band excess.}
\tablenotetext{8}{Spectral type estimated from $\Delta$V between $\sigma$~Ori~Aa and Ab.}
\tablenotetext{9}{This spectral type is from \citet{guetter81}.  We have included this star because we doubt the luminosity class (see note 5).  If the luminosity class is correct, this star is unlikely to be a member of the cluster.}
\tablenotetext{a}{Spectral types from \citet{edwards76}}
\tablenotetext{b}{Reddening free colors are from \citet{kh95} except where otherwise noted.}
%\tablenotetext{b}{This V magnitude is from \citet{paper3}.}
\tablenotetext{c}{Spectral types from \citet{schild71}}
%\tablenotetext{d}{Data from \citet{gw58}.}
%\tablenotetext{f} {Spectra from NOTHING!!}
\tablenotetext{g}{Data from \citet{guetter81}.}
\tablenotetext{h}{Data from Table 12 of \citet{sk82}.}
%\tablenotetext{i}{See Section~\ref{hd294279}}
%{\citet{cab_thesis} report a spectral type of F3 to F5 and the detection of Li in the spectrum.  If the spectral type were as late as F5, then HD~294279 would lie near the isochrone, but its observed B$-$V is 0.39 which is much too blue for an F5 star with the clusters measured E(B$-$V).  Such a star should have B$-$V=0.50. }
%\tablenotetext{n}{These stars do not have published MK spectral types.  SIMBAD lists spectral types from the second extension of the HD catalog.  See \citet{hdext95} and references therein.  The values in Table~\ref{true_table} are from our spectrascopic data. See Section~\ref{analysis}.
%\tablenotetext{s}{Data from SIMBAD.}
\end{deluxetable}

\begin{deluxetable}{lcccccc}
\tabletypesize{\scriptsize}
%\rotate
\tablecaption{Distance Estimates (pc) \label{dist_table}}
\tablewidth{0pt}
\tablehead{
\colhead{ZAMS/Color} & \colhead{$\frac{Fe}{H}$=+0.04} & \colhead{$\frac{Fe}{H}$=0.0} & \colhead{$\frac{Fe}{H}$=$-$0.05} & \colhead{$\frac{Fe}{H}$=$-$0.16} &
\colhead{$\frac{Fe}{H}$=$-$0.27} & \colhead{Color Correction}
}
\startdata
Turner B$-$V        & 442$\pm$20        & 436$\pm$20        & 428$\pm$19        & 412$\pm$19        & 397$\pm$18        & Mean E(B$-$V) \\
Schmidt-Kaler B$-$V & 462$_{-35}^{+14}$ & 456$_{-35}^{+14}$ & 448$_{-34}^{+14}$ & 431$_{-33}^{+13}$ & 415$_{-31}^{+13}$ & Mean E(B$-$V) \\
Schmidt-Kaler U$-$B & 488$_{-20}^{+18}$ & 481$_{-20}^{+18}$ & 473$_{-19}^{+17}$ & 455$_{-19}^{+17}$ & 438$_{-18}^{+16}$ & Mean E(B$-$V)\tablenotemark{1} \\
Schmidt-Kaler U$-$B & 458$_{-17}^{+15}$ & 452$_{-17}^{+15}$ & 444$_{-16}^{+15}$ & 427$_{-16}^{+14}$ & 411$_{-15}^{+13}$ & Mean E(U$-$B) \\
Schmidt-Kaler U$-$B & 452$\pm$14        & 445$\pm$14        & 438$\pm$14        & 422$\pm$13        & 406$\pm$13        & SpT\tablenotemark{2} \\
Schmidt-Kaler U$-$V & 492$_{-31}^{+18}$ & 485$_{-31}^{+18}$ & 477$_{-30}^{+17}$ & 459$_{-29}^{+17}$ & 442$_{-28}^{+16}$ & Mean E(B$-$V)\tablenotemark{1} \\
Schmidt-Kaler U$-$V & 457$_{-16}^{+28}$ & 451$_{-16}^{+28}$ & 443$_{-16}^{+27}$ & 426$_{-15}^{+26}$ & 411$_{-14}^{+25}$ & Mean E(U$-$V) \\
Schmidt-Kaler U$-$V & 447$_{-7}^{+13}$  & 441$_{-7}^{+13}$  & 433$_{-7}^{+13}$  & 417$_{-7}^{+12}$  & 402$_{-6}^{+12}$  & SpT\tablenotemark{2} \\
\hline
Combined Estimate\tablenotemark{3}   & 450$\pm$20 & 444$\pm$20 & 436$\pm$20     & 420$\pm$20        & 404$\pm$20        & \\

\enddata
% Text for table notes should follow after the \enddata but before
% the \end{deluxetable}. Make sure there is at least one \tablenotemark
% in the table for each \tablenotetext.
\tablenotetext{1}{Assuming a standard reddening law with E(U$-$B)=0.72E(B$-$V).}
\tablenotetext{2}{We adopted the colors for each star's spectral type given by \citet{kh95}.}
\tablenotetext{3}{The estimated uncertainties include only the random error from the fit.  The systematic uncertainty from the choice of ZAMS is at least 10~pc.  Combined with the uncertainty on the metallicity, the total systematic error is $\sim$25~pc.}

\end{deluxetable}

%%% Tables may also be prepared as separate files. See the accompanying
%%% sample file table.tex for an example of an external table file.
%%% To include an external file in your main document, use the \input
%% command. Uncomment the line below to include table.tex in this
%% sample file. (Note that you will need to comment out the \documentclass,
%% \begin{document}, and \end{document} commands from table.tex if you want
%% to include it in this document.)


%% The following command ends your manuscript. LaTeX will ignore any text

\end{document}